\documentclass[article]{jfm}

\usepackage{graphicx}

\usepackage{newtxtext}
\usepackage{newtxmath}
\usepackage{natbib}
\usepackage{hyperref}
\hypersetup{
    colorlinks = true,
    urlcolor   = blue,
    citecolor  = black,
}

\newcommand{\RomanNumeralCaps}[1]

\title{Statistics of rogue waves in isotropic wave fields}

\author{Guillaume Michel\aff{1}
\corresp{\email{guillaume.michel@upmc.fr}}, F\'elicien Bonnefoy\aff{2}, Guillaume Ducrozet\aff{2} \and   Eric Falcon\aff{3}}

 \affiliation{\aff{1}Sorbonne Université, CNRS, Institut Jean Le Rond d’Alembert, F-75005 Paris, France\aff{2}\'Ecole Centrale de Nantes, LHEEA, UMR 6598 CNRS, F-44321 Nantes, France\aff{3}Université Paris Cité, CNRS, MSC, UMR 7057, F-75013 Paris, France}

\begin{document}

\maketitle

\begin{abstract}
We investigate the statistics of rogue waves occurring in the inverse cascade of surface gravity wave turbulence. In such statistically homogeneous, stationary and isotropic wave fields, low-frequency waves are generated by nonlinear interactions rather than directly forced by a wave maker. This provides a laboratory realization of arguably the simplest nonlinear sea state, in which long-time acquisitions are performed and compared with theoretical models. The analysis of thousands of rogue waves reveals that some of their properties crucially depend on four-wave resonant interactions, large crests being for instance more likely than predicted by second-order models.
\end{abstract}

\section{Introduction}
\label{sec:intro}

As a result of cheaper computational storage and improved sensors, the number of surface waves included in databases of field measurements has soared over recent decades, going from fifty thousand at the end of the 1970s to hundreds of millions in 2020 \citep{Forristall_1978, Karmpadakis_2020}. They allow for systematic correlation studies with hindcast data, evidencing, for instance, that the probability of occurrence of rogue waves (RWs) is independent of the instantaneous wind speed and direction \citep{Christou_2014}. These approaches are undoubtedly valuable as they single out the environmental conditions that favour the occurrence of RW but remain far from being exhaustive. For instance, the overwhelming majority of deep-water waves discussed in this context in \cite{Christou_2014} share the same directions of swell and current, precluding the possible evidence of generation of RW by wave-current interactions, a phenomenon yet recognized as a promising outlook \citep{Adcock_2014, Toffoli_2015, DUCROZET_2021}. More fundamentally, drawing a comprehensive theory of RWs based on these results is complicated by the lack of statistically stationary states: in practice, wave elevation time series from different storms are spliced into 20 min samples then recombined with others sharing similar proxies (e.g., wave mean frequency, mean direction of propagation, etc.), which unavoidably introduces a bias and explains why the distribution of rare events such as RWs is still discussed.
 
To assess theoretical models, laboratory experiments nicely complement field experiments since they provide long-time statistics under controlled conditions. Most of them take place in long flumes in which \textit{unidirectional} waves, also referred to as ``long-crested waves'', are randomly generated by a wave maker and propagate over more than a hundred meter before being damped by a beach. Such experiments typically report a transient overshoot of the kurtosis, of the spectral width and of the RW probability associated with the emergence of high-amplitude structures locally akin to the so-called Peregrine soliton (PS) \citep{Onorato_2004, Onorato_2005, Onorato_2006, Shemer_2009, Shemer_2010a, Shemer_2010b,Cazaubiel_2018,Dematteis_2019, Michel_2020}. This dynamics can be modelled by the nonlinear Schrödinger equation (NLSE), an exact solution of the latter, localized in both space and time, being the PS. The instability of a continuous wave train, called the ``modulation instability'' and generating RWs, can also be studied in long one-dimensional flumes and described by the NLSE, see, e.g.,  \citet{Lighthill_1965, Benjamin_1967,Lake_1977, Melville_1982, Chabchoub_2017} and references therein. All these results strongly depend on the directionality of the wave field, as shown both theoretically through the existence of transverse instabilities \citep{Badulin_2012, Ablowitz_2021}, numerically \citep{Onorato_2002, Soquet_2005, Gramstad_2007,Toffoli_2008} and experimentally \citep{Waseda_2006, Onorato_2009}, questioning their relevance in accounting for \textit{in situ} RWs.

On the other hand, another set of experiments investigate the theory of weak wave turbulence (WWT), which predicts how energy spreads among random waves in nonlinear interaction \citep{Falcon_2022}. They take place in basins with reflecting walls and deal with \textit{isotropic} or at least strongly multidirectional waves (``short-crested waves'').  Until recently, they essentially consisted of generating waves with a wavelength a fraction of the length of the basin and measuring the energy cascade toward small scales \citep{Denissenko_2007, Lukaschuk_2009, Nazarenko_2010, Deike_2015, Aubourg_2017, Campagne_2018}. A breakthrough occurred in 2020, when it was evidenced that forcing multidirectional random waves of short wavelengths in such basins not only generates even shorter wavelengths but also larger ones, corresponding to the inverse cascade of WWT \citep{Falcon_2020}. Such wave fields are valuable for the study of RWs since the waves involved in their dynamics are spontaneously generated by nonlinear interactions rather than directly forced by the wave-maker. Moreover, they verify isotropy, homogeneity and stationarity, and therefore offer a unique framework to confront theoretical predictions on RWs to a simplified though strongly nonlinear model of the sea state.  The present study reports the statistics of thousands of RWs measured in such a state and investigates the effect of high-order nonlinearities.

\section{Experimental setup}

\begin{figure}
    \begin{center} \includegraphics[width=7.5cm]{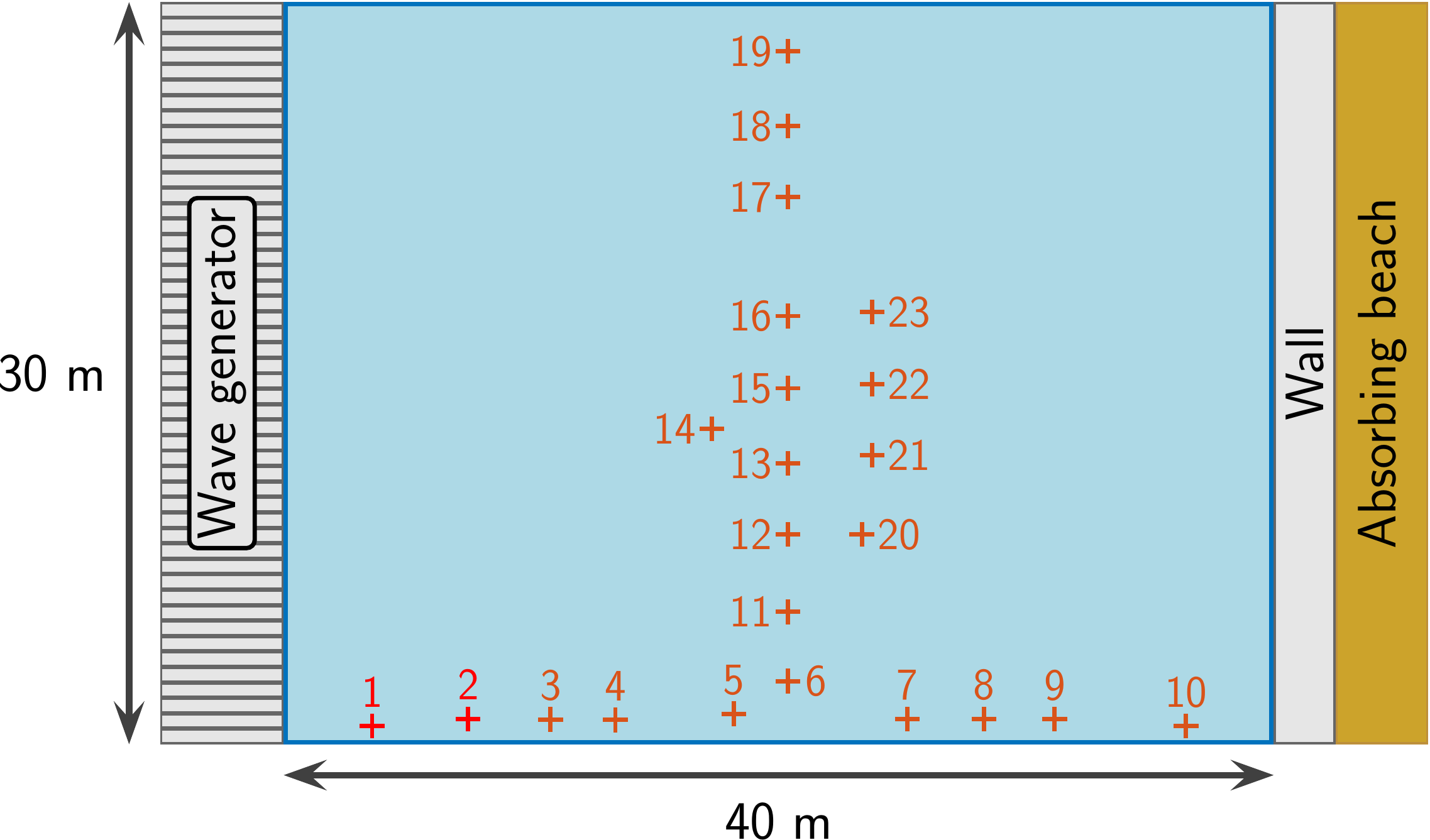} \hspace{2mm} \includegraphics[width=5.2cm]{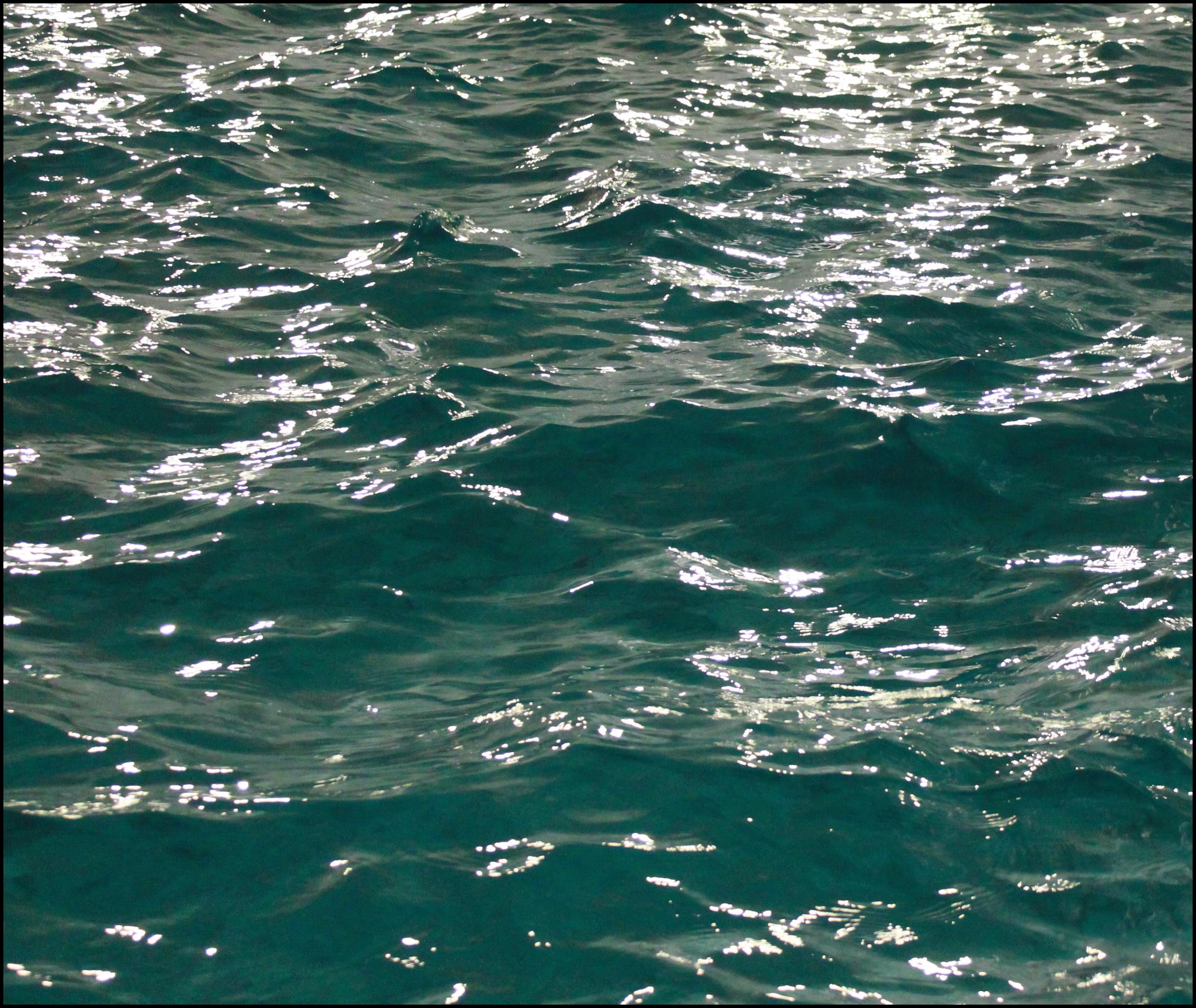}
    \caption{Left: Experimental setup showing the 48 flap wave generator, the end wall and the 23 probes. Probes 1 and 2 are used to verify the wave maker behaviour and are not included in the data analysis. Right: Photograph of a typical wave field (Run 3, the horizontal field of view is approximately one metre).}
    \label{bassin}
\end{center}
\end{figure}

Experiments are carried out in the large-scale basin (40 m long $\times$ 30 m wide $\times$ 5 m deep) of Ecole Centrale de Nantes, France. At one end of the basin, 48 flaps of width $\ell = 0.62~\mathrm{m}$ are driven independently by different realizations of white noise filtered in the $[f_0-\Delta f, f_0+\Delta f]$ frequency range, with $f_0=1.8~\mathrm{Hz}$ the central frequency and $\Delta f = 0.2~\mathrm{Hz}$ the bandwidth. Therefore, each flap generates independent waves of frequency around $f_0$ (wavelength $\lambda_0 = 0.48~\mathrm{m}$, group velocity $v_g=0.43~\mathrm{m}\cdot \mathrm{s}^{-1}$) with a directional spread that can be estimated as $\theta = 2\times (\lambda_0/\ell)  = 88^\mathrm{o}$. Three forcing amplitudes are considered, hereafter referred to, in increasing order, as Runs 1 to 3. At the other end, a solid vertical wall is built ahead of the usual beach. This setup is sketched in Fig. \ref{bassin} Left.

As reported in \citet{Falcon_2020}, a statistically stationary, homogeneous and isotropic nonlinear steady state is reached after a transient of up to twenty minutes. The general picture is as follows: during this transient, the waves generated at $f_0$ by the flaps travel over nearly 70 times the length of the basin ($20~\mathrm{min}/v_g = 2.8~\mathrm{km}$). As they propagate, nonlinear effects such as four-wave resonant interactions and very steep structures spread energy in all directions. Some of these strongly nonlinear effects visible from the shore are found to occur homogeneously in the basin, e.g. capillary waves generated  by large gravity waves. Note that white capping is not observed, see Fig. \ref{bassin} Right.

\begin{figure}
    \begin{center} \includegraphics[width=8cm]{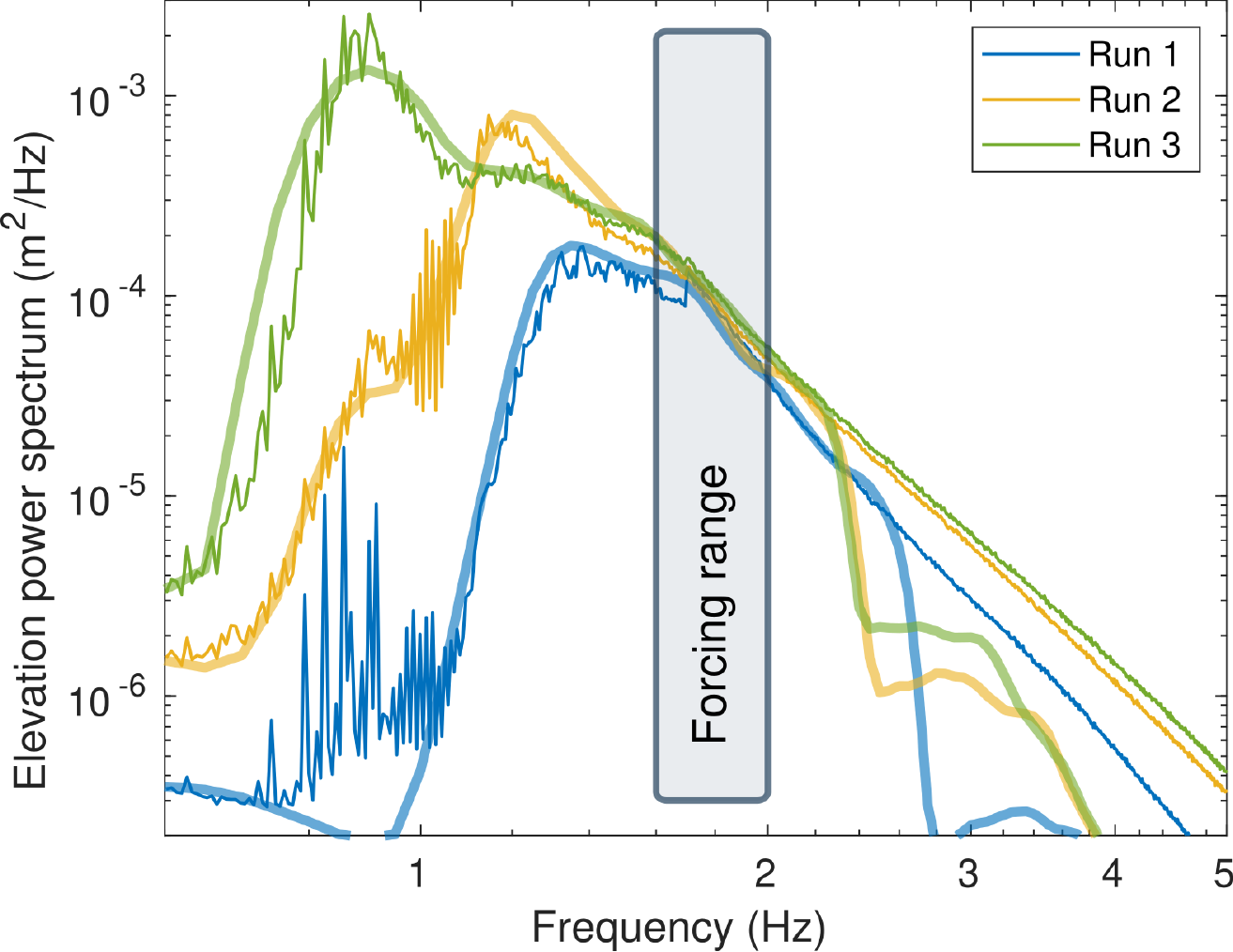} \end{center}

    \caption{PSD of surface elevation for the three different steady states considered. Thin lines correspond to experimental results and thick ones to their numerical model. The forcing bandwidth is also displayed.}
   \label{PSD}
 \end{figure}

The surface elevations $\left\lbrace\eta_i(t)\right\rbrace_{i=1 \dots 23}$ are recorded by 23 resistive probes of vertical resolution 0.1 mm and frequency resolution 100 Hz during 27 to 30 hours depending on the run. These measurements can be used to verify the claims of stationarity, homogeneity and isotropy. Stationarity is confirmed through the time evolution of statistical measurements of the wave field, e.g. the standard deviation of surface elevations computed over one minute samples, and is achieved after up to twenty minutes, see figures in \cite{Falcon_2020}. The transients are not investigated in this study and only measurements performed in the steady-state regimes are hereafter discussed. All probes are found to measure a similar standard deviation of surface elevation up to a maximum relative difference of $10\%$: homogeneity is closely achieved, and to remove the small remaining bias each signal is normalized by the standard deviation of the corresponding probe. Isotropy is the most challenging assumption to test since it cannot be investigated from a single elevation signal. The cross-correlation between pairs of elevation signals is therefore introduced. For each run, it is computed as
\begin{equation}
R_{i,j}(\tau) = \frac{\langle \eta_i(t) \eta_j(t+\tau) \rangle}{\sqrt{\langle \eta_i(t)^2 \rangle \langle \eta_j(t)^2 \rangle}},
\end{equation}
where $\langle \cdot \rangle$ denotes a temporal averaging. Over all runs,  all lags $\tau$ and all probes $i \neq j$, $\vert R_{i,j} \vert $ remains less than 0.2 and the probes are therefore largely uncorrelated, as expected from their large spatial separation. Nevertheless, the remaining correlations evidence that $R_{i,j}(\tau)$ is almost symmetric, i.e. that the wave field is essentially isotropic ($R_{i,j}(\vert \tau \vert)$ and $R_{i,j}(-\vert \tau \vert)$, respectively, account for signals propagating from $i$ to $j$ and from $j$ to $i$). Quantitatively, with $(i \neq j) \in [13,14,15]$ standing for the three close central probes and $\tau_\mathrm{max}$ such that $R_{i,j}(\tau_\mathrm{max})$ is maximum,
\begin{equation}
\left \vert \frac{R_{i,j}(\tau_\mathrm{max}) - R_{i,j}(-\tau_\mathrm{max})}{R_{i,j}(\tau_\mathrm{max})} \right \vert  < 0.16,
\end{equation}
a strong indication toward isotropy. Finally, note that the power spectrum density $S_\eta(f)$ (PSD),  reported in Fig. \ref{PSD}, reveals that most of the energy is located at frequencies \textit{smaller than} the forcing range, corresponding to waves forced by nonlinear interactions. These PSDs present some features theoretically predicted for the inverse cascade of WWT and derived under the assumption of stationarity, homogeneity and isotropy \citep{Falcon_2020}.

It is instructive to detail the energy budget of this wave field. Energy is injected in the wave system at a rate $\mathcal{P}_\mathrm{inj}$ that can be measured through decay experiments and is of several watts (see \cite{Falcon_2020}, note that this power is much smaller than the one supplied to the wave maker). Conversely, the power dissipated by viscosity at high frequency (> 2Hz) at the surface boundary layer can be estimated from the experimental PSD and reads \citep{Miles_1967}
 \begin{equation}
 \mathcal{P}_\mathrm{diss} = 2 S \rho g \int_{2~\mathrm{Hz}}^{\infty} S_\eta(f) \alpha(f) \mathrm{d}f,
 \end{equation}
with $S= 30\times 40 ~\mathrm{m}^2$ the surface of the basin, $\rho=10^3~\mathrm{kg}\cdot \mathrm{m}^{-3}$ the density, $g=9.81~\mathrm{m}\cdot \mathrm{s}^{-2}$ the acceleration due to gravity, $\alpha(f) = 2\nu k^2 = 2 \nu (2\pi f)^4/g^2$ the damping coefficient for clean water and $\nu = 10^{-6}\mathrm{m}^2\cdot \mathrm{s}^{-1}$ the kinematic viscosity. We find typically $\mathcal{P}_\mathrm{diss} \sim \mathcal{P}_\mathrm{inj}/10$, meaning that most of the energy is dissipated by another mechanism than viscous dissipation of high-frequency waves in the bulk. We believe that this mechanism is linked with the nonlinear dynamics at large scales, which involves very steep structures acting as localized sources of dissipation (e.g., cusps of very steep slope).

\section{Numerical model}

To identify high-order nonlinear effects in the experimental data, these wave fields are reproduced numerically up to second-order nonlinearities. The elevation at a given location is computed as $\eta(t) = \eta^{(1)}+ \eta^{(2)}$, where the linear contribution $\eta^{(1)}$ is the sum of $N_\omega \times  N_\theta=512$ independent progressive waves ($N_\omega=16$ angular frequencies, each of them associated with $N_\theta=32$ directions), and $\eta^{(2)}$ is the nonlinear correction. More precisely, $\eta^{(1)}$ reads
\begin{equation}
\eta^{(1)}(t) = \sum_{n_\omega=1}^{N_\omega} \sum_{n_\theta=1}^{N_\theta} a_{n_\omega,n_\theta} \cos \left( - \omega_{n_\omega} t + \phi_{n_\omega,n_\theta}\right),\label{num_1}
\end{equation}
where $a_{n_\omega,n_\theta}$ are random numbers drawn from normal distributions of zero mean and standard deviations $A_{n_\omega}$. The phase constants $\phi_{n_\omega,n_\theta}$ are uniformly distributed in the range $[0, 2\pi]$. The leading-order nonlinear correction $\eta^{(2)}$ stems from \citet{LH_1963} (up to a correction factor of one half, see \citet{Srokosz_1986}). In particular, it involves the wave vectors of the linear waves, set to model an isotropic wave field as
\begin{equation}
\mathbf{k}_{n_\omega,n_\theta} =  \frac{\omega_{n_\omega}^2}{g} \left[\cos \left(\frac{2\pi n_\theta}{N_{\theta}}\right) \mathbf{e}_x+ \sin\left(\frac{2\pi n_\theta}{N_{\theta}}\right) \mathbf{e}_y \right].
\end{equation}
The angular frequencies $\left\lbrace \omega_{n_\omega} \right\rbrace$ are linearly distributed in a given range with $\Delta \omega = 2\pi \times 0.1~\mathrm{rad}\cdot \mathrm{s}^{-1}$. Both this range and the constants $\left\lbrace A_{n_\omega} \right\rbrace$ are adjusted to reproduce the experimental spectra at large scale, see Fig. \ref{PSD}. For each run, $5\times 10^7$ values of $\eta(t=0)$ and millions of waves from time series of $\eta(t)$ with a time step of $0.01 ~\mathrm{s}$ are computed from independent drawings of $\left\lbrace a_{n_\omega,n_\theta},\phi_{n_\omega,n_\theta} \right\rbrace$. The former are used to obtain the data reported in Tab. \ref{Table} and Fig. \ref{PSD}-\ref{PDF} whereas waves are documented in Fig. \ref{troughs_crests} - \ref{Large_waves}. 

\begin{table}
\begin{center}\def~{\hphantom{0}}
\begin{tabular}{c|c|c|c|c|c|c|c|c|c|}
 & \multicolumn{3}{c|}{Run 1}& \multicolumn{3}{c|}{Run 2} & \multicolumn{3}{c|}{Run 3}\\
  & exp&num&th& exp&num&th & exp&num&th\\
$\sigma$ (cm) & 0.98 &1.02& & 1.63 &1.74& & 2.31 &2.37& \\
$\mathcal{S}$ & 0.21 & 0.19& 0.20 & 0.25 &0.24& 0.23 & 0.22 &0.20& 0.22 \\
$K-3$& 0.18 &0.06 & 0.06 & 0.20 &0.09& 0.07 & 0.17 &0.07& 0.04 \\
$\nu$& 0.25 &0.21&  & 0.30 &0.22& & 0.36 &0.31& \\
$f_0$ (Hz)& 1.68 &1.60& & 1.43 &1.37& & 1.18 &1.13& \\
$f_\mathrm{p}$ (Hz) & 1.38 &1.35& & 1.15 &1.20& & 0.90 &0.90& \\
$f_T$ (Hz) & 1.59 &1.54& & 1.33 &1.31& & 1.06 &1.04& \\
$\varepsilon_T$ & 0.10 &0.10& & 0.12 &0.12& & 0.11 &0.10& \\
$N_\mathrm{tot}$ & \multicolumn{2}{c|}{$3~779~963$} & & \multicolumn{2}{c|}{$3~385~889$} & & \multicolumn{2}{c|}{$2~548~368$}& \\
$N_\mathrm{RW}$ & 937 &840&& 899 &798& & 475 &450& \\
\end{tabular}
\end{center}
\caption{Standard deviation $\sigma$, skewness $\mathcal{S}$, kurtosis $K$, dimensionless spectral bandwidth $\nu$, mean frequency $f_0$, peak frequency $f_\mathrm{p}$, Tayfun frequency $f_T$ and steepness $\varepsilon_T$ based on $f_T$. ``exp'' denotes experimental measurements, ``num'' numerical models and ``th'' theoretical estimates given by Eq. \eqref{S_th} and \eqref{C_th} and computed based on the experimental PSD. The number of waves $N_\mathrm{tot}$ and rogue waves $N_\mathrm{RW}$, defined as $H>2 H_S$ with $H_S = 4 \sigma$, are also reported.}
\label{Table}
\end{table}

\section{Moments}

The first moments of $\eta(t)$ from experiments and numerical models are reported in Tab. \ref{Table}. The standard deviation $\sigma = \langle \eta ^2 \rangle^{1/2}$ is found to increase with the forcing amplitude, while the skewness $\mathcal{S} = \langle \eta^3 \rangle / \sigma^3$ and the kurtosis $K  = \langle \eta^4 \rangle / \sigma^4$ remain roughly constant. Other characteristics of sea states are computed, namely the dimensionless spectral bandwidth $\nu = (m_0m_2/m_1^2-1)^{1/2}$, with $m_n = \int S_\eta(f) f^n \mathrm{d}f$ the spectral moments, the mean frequency $f_0 = m_1/m_0$, the peak frequency $f_\mathrm{p}$, the Tayfun frequency $f_T = f_0/[1+\nu^2(1+\nu^2)^{-3/2}]$ discussed later in the manuscript \citep{Tayfun_1993, Tayfun_2007} and the steepness $\varepsilon_T = (2 \pi f_T)^2 \sigma/g$ based on $f_T$, with $g$ the acceleration due to gravity. The dimensionless parameters measured experimentally ($\mathcal{S}$, $K$, $\nu$ and $\varepsilon_T$) correspond to typical values observed in the ocean, although field measurements yield $f_{0,\mathrm{p},T}= O(0.1)~\mathrm{Hz}$ and $\sigma = O(1)~\mathrm{m}$ \citep{Christou_2014}. This confirms that the wave field under study shares the complex dynamics at work in the ocean while allowing the recording of ten times more waves over the same acquisition time. 

The skewness $\mathcal{S}$ can be compared with theoretical predictions. The linear model reduces surface elevation to a sum of independent progressive waves of various frequencies and amplitudes ($\eta^{(1)}(t)$ in \eqref{num_1}), for which $\mathcal{S}$ vanishes. In the 1960s, Longuet-Higgins computed the second-order nonlinear correction $\eta^{(2)}(t)$ and showed that it only involves non-resonant interactions, mathematically of the form of progressive waves that do not verify the linear dispersion relation, the so-called ``bound waves'' \citep{LH_1963}. The skewness then becomes non-zero and can be inferred from $S_\eta(f)$:  simplified under the assumption of an isotropic wave field, it reads
\begin{align}\label{S_th}
\mathcal{S} &=  \int \frac{3k_1}{2\pi \sigma^3} S_\eta(k_1) S_\eta(k_2) I\left( \frac{k_2}{k_1} \right) \mathrm{d}k_{1,2}
\end{align}
where $I$ is an explicit function, see Appendix \ref{appB}. Further, assuming a narrowband frequency spectrum ($\nu \ll 1$, i.e., $f_0=f_T=f_\mathrm{p}$) numerically yields $\mathcal{S} = 2.07 \varepsilon_T$, in contrast to $\mathcal{S} = 3 \varepsilon_T$ for unidirectional waves of a narrowband frequency spectrum. The theoretical prediction of $\mathcal{S}$ computed from Eq. \eqref{S_th} together with the experimental PSD $S_\eta$ is reported in Tab. \ref{Table}: it accounts  for both numerical models and experimental results.

Several decades later, Janssen built on the canonical transformation introduced in \cite{Zakharov_1968} to derive the surface elevation up to the next order and to consistently compute the deviation of the kurtosis from three \citep{Janssen_2009}. Disentangling resonant and non-resonant interactions, he obtained
\begin{equation}
K=3(1+C_4^\mathrm{dyn}+C_4^\mathrm{can}),
\end{equation}
where $C_4^\mathrm{dyn}$ results from four-wave resonant interactions and only allows analytic expressions for spectra that are narrow in frequency and direction \citep{Fedele_2015, Janssen_2019}.  In contrast, $C_4^\mathrm{can}$ is associated with bound waves and can be inferred directly from $S_\eta(f)$: for an isotropic wave field,
\begin{align}
C_4^\mathrm{can}&=\int \frac{k_1^2}{\pi^2\sigma^4} S_\eta(k_1) S_\eta(k_2) S_\eta(k_3) \psi \left( \frac{k_2}{k_1}, \frac{k_3}{k_1}\right) \mathrm{d}k_{1,2,3}~,\label{C_th}
\end{align}
where $\psi$ is another explicit function, see Appendix \ref{appC}. Furthermore, if the spectrum is narrowband in frequency, it reduces to $C_4^\mathrm{can} = 2.75\varepsilon_T^2$. The theoretical values of $3C_4^\mathrm{can}$ computed from Eq. \eqref{C_th} and the experimental PSD $S_\eta$ are reported in Tab. \ref{Table}. They match our numerical models, in which no resonant interaction takes place, but strongly differs from experimental measurements. This demonstrates that fou- wave interactions not only generate the low-frequency waves under study but also crucially affect their statistics. Note that a similar conclusion has been reached in a regime of capillary wave turbulence dominated by four-wave interactions \citep{Xia_2010, Shats_2010}.

\section{Probability Density Functions (p.d.f.s)}

\begin{figure}
    \begin{center} \includegraphics[height=5.0cm]{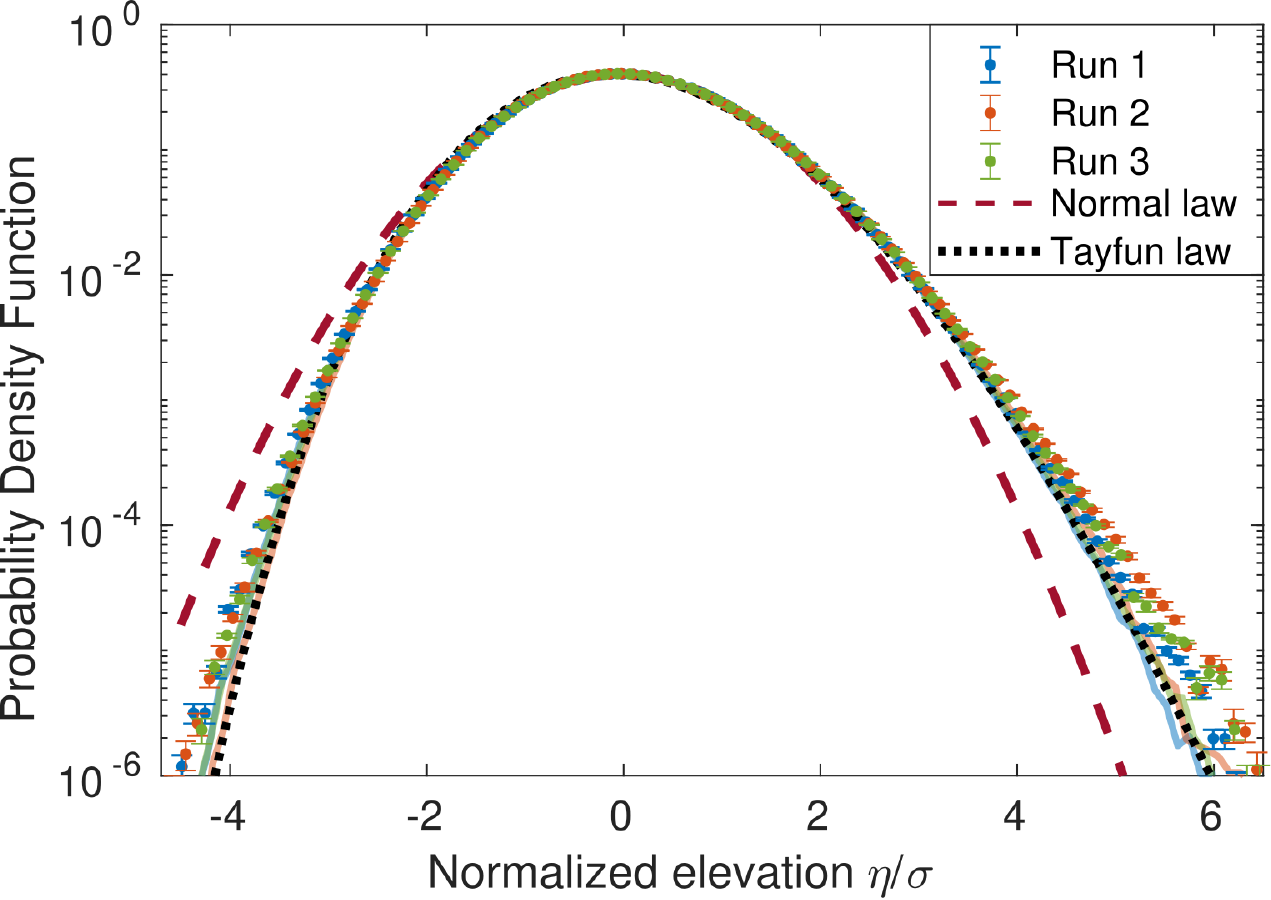} \includegraphics[height=4.86cm]{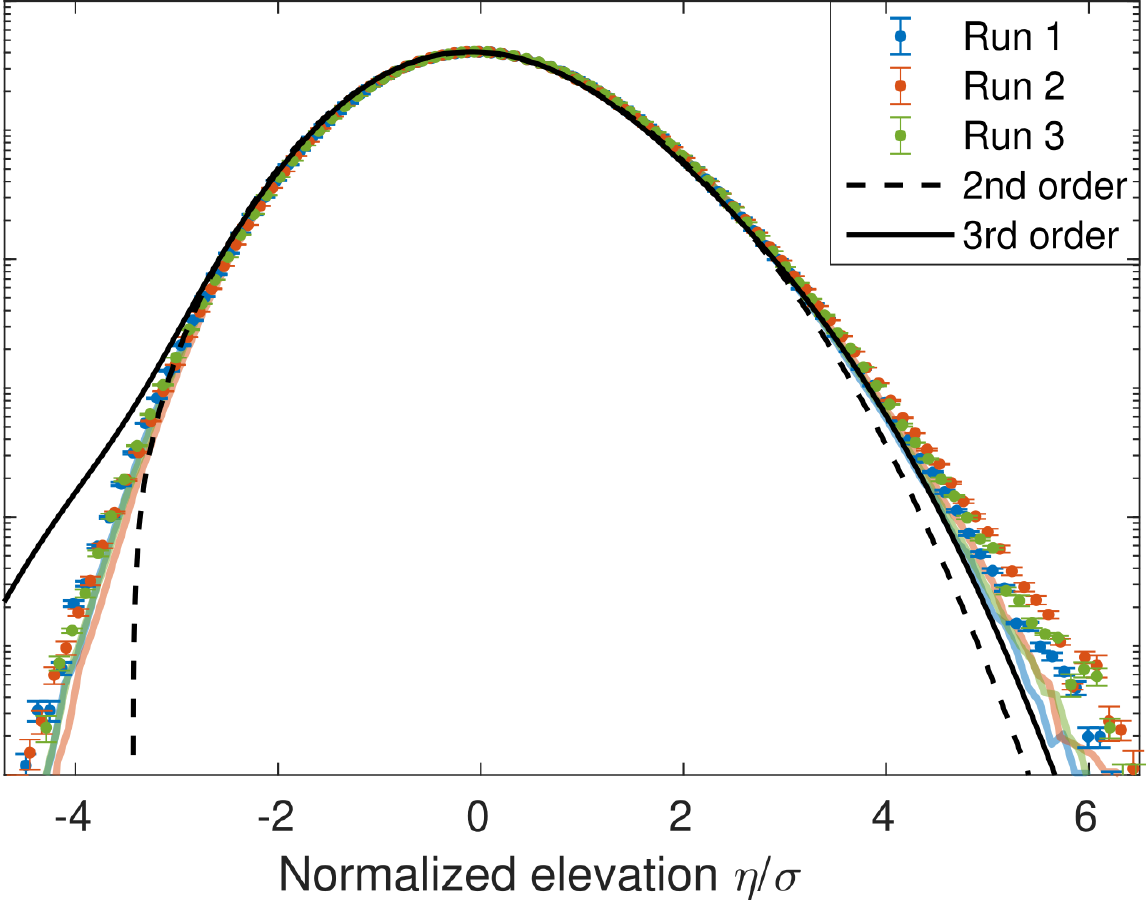}  \end{center}
\caption{The p.d.f.s of the normalized surface elevation $\eta/\sigma$ from experiments (symbols) and numerical models (thick coloured lines), compared with (left) a normal law and a Tayfun law and (right) the second- and third-order Gram-Charlier series computed with $\mathcal{S} = K-3 = 0.2$.}
\label{PDF}
\end{figure}
 
The p.d.f.s of experimental and numerical normalized surface elevations $f(u=\eta/\sigma)$ are reported in Fig. \ref{PDF}, along with a normal law of zero mean and unit variance, a Tayfun law and two Gram-Charlier series. The normal distribution describes linear waves and accounts neither for the finite skewness nor for a kurtosis other than three. The Tayfun law corresponds to \textit{unidirectional} and narrowband waves with second-order nonlinearities \citep{Tayfun_1980}, see Appendix \ref{appD} for its analytic expression. It only depends on the steepness $\varepsilon_T$ and has been shown empirically to provide a fair estimate of the \textit{tails} of $f(u)$ for isotropic and broadbanded waves as well, provided that $\varepsilon_T$ is artificially tuned to generate the observed skewness (0.24 in the case of Fig. \ref{PDF} (left)) \citep{Aubourg_2017, Falcon_2020}. It is found here to fit the tails of the numerical p.d.f.s and to underestimate the experimental ones. This difference in the probability of extreme surface elevations translates into the difference in kurtosis discussed before. The p.d.f.s are also compared with the low-order Gram-Charlier series commonly used in theoretical work on surface waves, see Appendix \ref{appD2} for definitions. They are reported in Fig. \ref{PDF} (right) based on  the typical experimental values $\mathcal{S} = K-3 = 0.2$ from Tab.\ref{Table}. As observed in \cite{Klahn_2021}, they both underestimate large positive values and fail to capture large negative ones (for which the p.d.f. is either undefined, as for the second-order Gram-Charlier approximation, or largely above the experimental data, as for the third-order approximation).
 
Time series are then analysed in terms of zero down-crossing waves, i.e. events separated by zero crossings ($\eta = 0$) in which $\eta$ assumes negative then positive values \citep{Sea_state}. By definition, the wave height $H$ is the sum of the wave trough $\eta_\mathrm{T}$ (taken positive) and wave crest $\eta_\mathrm{C}$, the duration of the wave being the period $T$. In this manuscript, RWs are defined as waves for which $H > 2 H_S$, with $H_S = 4\sigma$ the significant wave height, whereas large crests are defined by $\eta_\mathrm{C} > 1.25 H_S$. The threshold $1.25 H_S=5\sigma$ corresponds to an alternative definition of RWs in the literature \citep{Fedele_2016}.  The numbers of recorded waves and RWs are reported in Tab. \ref{Table}.

\begin{figure}
    \begin{center} \includegraphics[width=7cm]{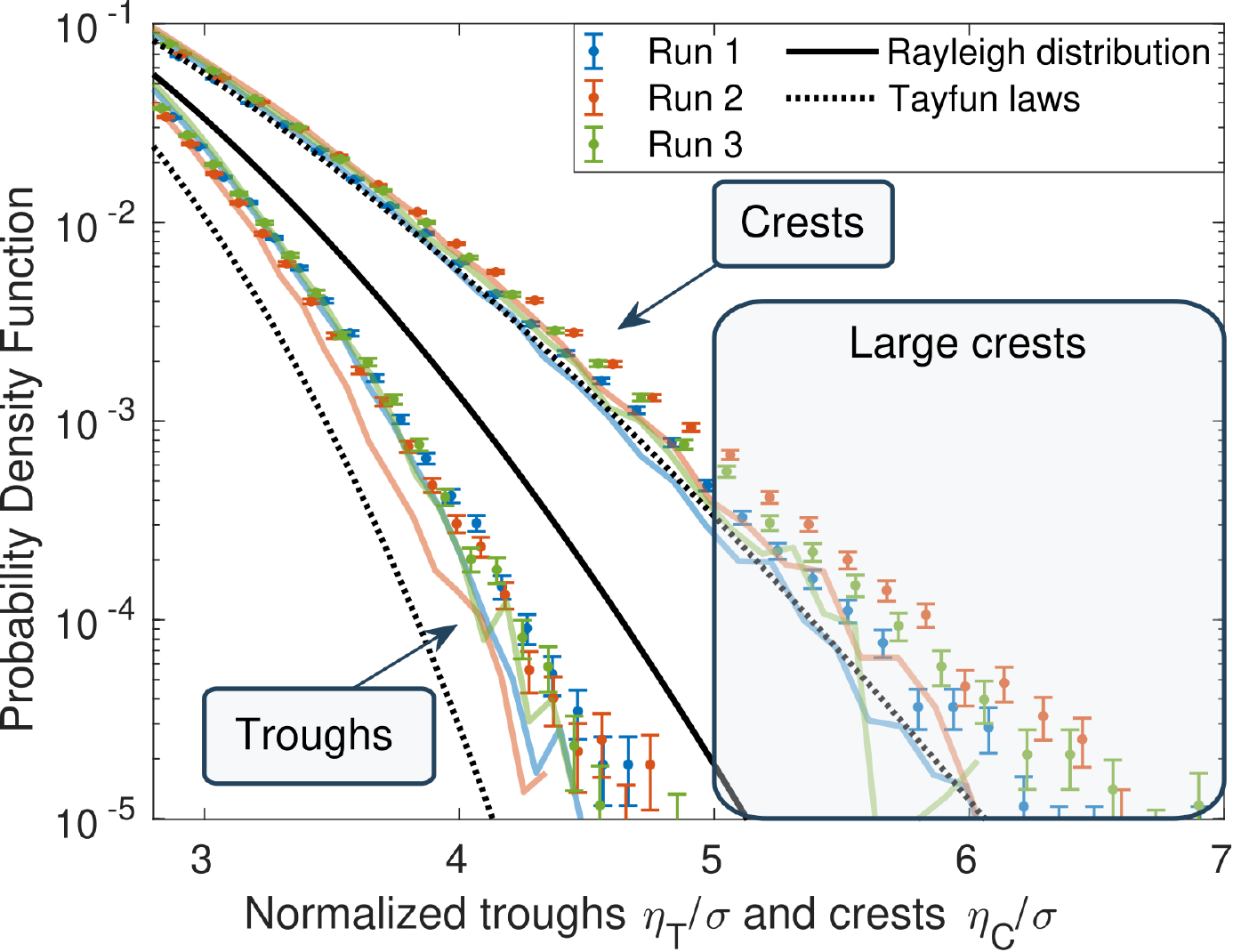} \end{center}
    
    \caption{The p.d.f.s  of the normalized wave troughs and crests compared with numerical models (thick coloured lines), a Rayleigh distribution (black solid line) and the first nonlinear corrections for unidirectional and narrowband waves (dotted lines).}
   \label{troughs_crests}
 \end{figure}

Consider first the p.d.f. of $\eta_\mathrm{C}$ and $\eta_\mathrm{T}$. For unidirectional and narrowband wave fields, they have been explicitly computed by Tayfun up to second-order nonlinearities \citep{Tayfun_1980}, see Appendix \ref{appE} for their analytic expressions. Similar to surface elevation, the p.d.f. of $\eta_\mathrm{C}$ has been empirically found  to fit the tails of multidirectional wave fields as well \citep{Soquet_2005, Denissenko_2007, Klahn_2021}. Both experimental and numerical p.d.f.s are reported in Fig. \ref{troughs_crests} along with the theoretical Rayleigh distribution ($f_\mathrm{R}(\xi) = \xi \exp(-\xi^2/2)$, capturing linear waves) and the Tayfun distributions with the steepness parameter tuned to describe a skewness of $0.24$. Our numerical models with bound waves only indicate that the fortuitous agreement between Tayfun's predictions for unidirectional waves and data from isotropic wave fields is restricted to crests. Moreover, one of the main outcomes of this work is that large crests are much more likely to be found experimentally than numerically or expected from the Tayfun law.

\begin{figure}
    \begin{center} \includegraphics[width=7cm]{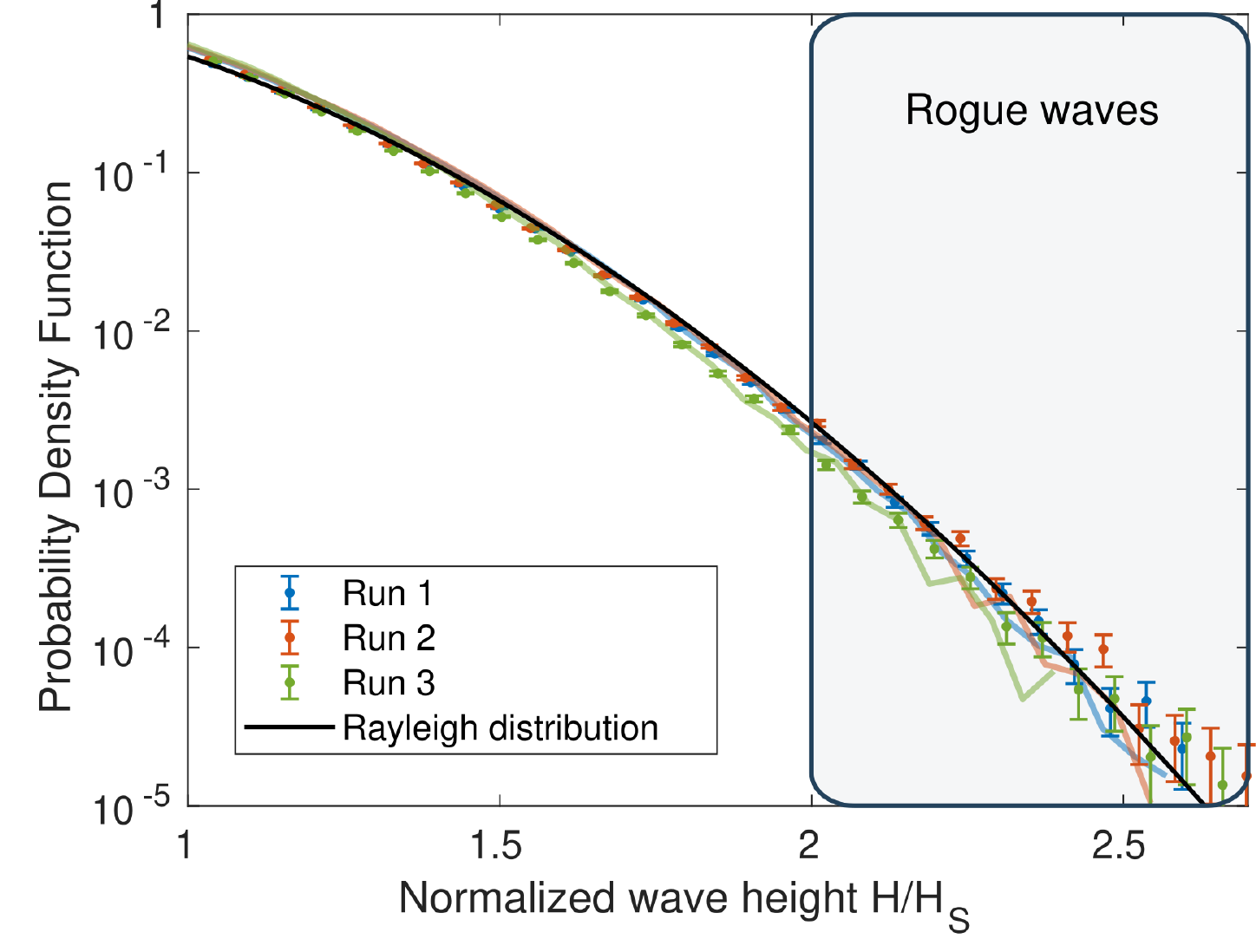} \end{center}
    \caption{The p.d.f.s of the normalized wave height $H/H_S$ from experiments (symbols) and numerical models (thick coloured lines), compared with a Rayleigh distribution.}
   \label{WH}
 \end{figure}

The wave height $H = \eta_\mathrm{C}+\eta_\mathrm{T}$ is then investigated.  As routinely observed, the distribution of $H/H_S$ as a function of the wave period $T$ peaks close to the inverse Tayfun frequency $f_T^{-1}$ \citep{Tayfun_1993, Tayfun_2007}, see the additional figures in Appendix \ref{appA}. The experimental and numerical p.d.f.s of $u=H/H_S$, reported in Fig. \ref{WH}, are compared with the Rayleigh distribution $f_R(u) = 4 u \exp(-2u^2)$, which describes narrowband waves with no assumption on directionality and is valid even when the second-order nonlinearities are included \citep{LH_1952, Tayfun_1980}. These data are all found to be similar. The wave height $H = \eta_\mathrm{C}+\eta_\mathrm{T}$ is therefore not only independent of second-order nonlinearities, as can be shown theoretically, but also seems to be largely independent of higher-order corrections. This is in sharp contrast with the statistics of $\eta_\mathrm{C}$ and $\eta_\mathrm{T}$ detailed above.

\section{Shape of large crests} \label{section_large_crests}

\begin{figure}
    \begin{center} \includegraphics[width=8cm]{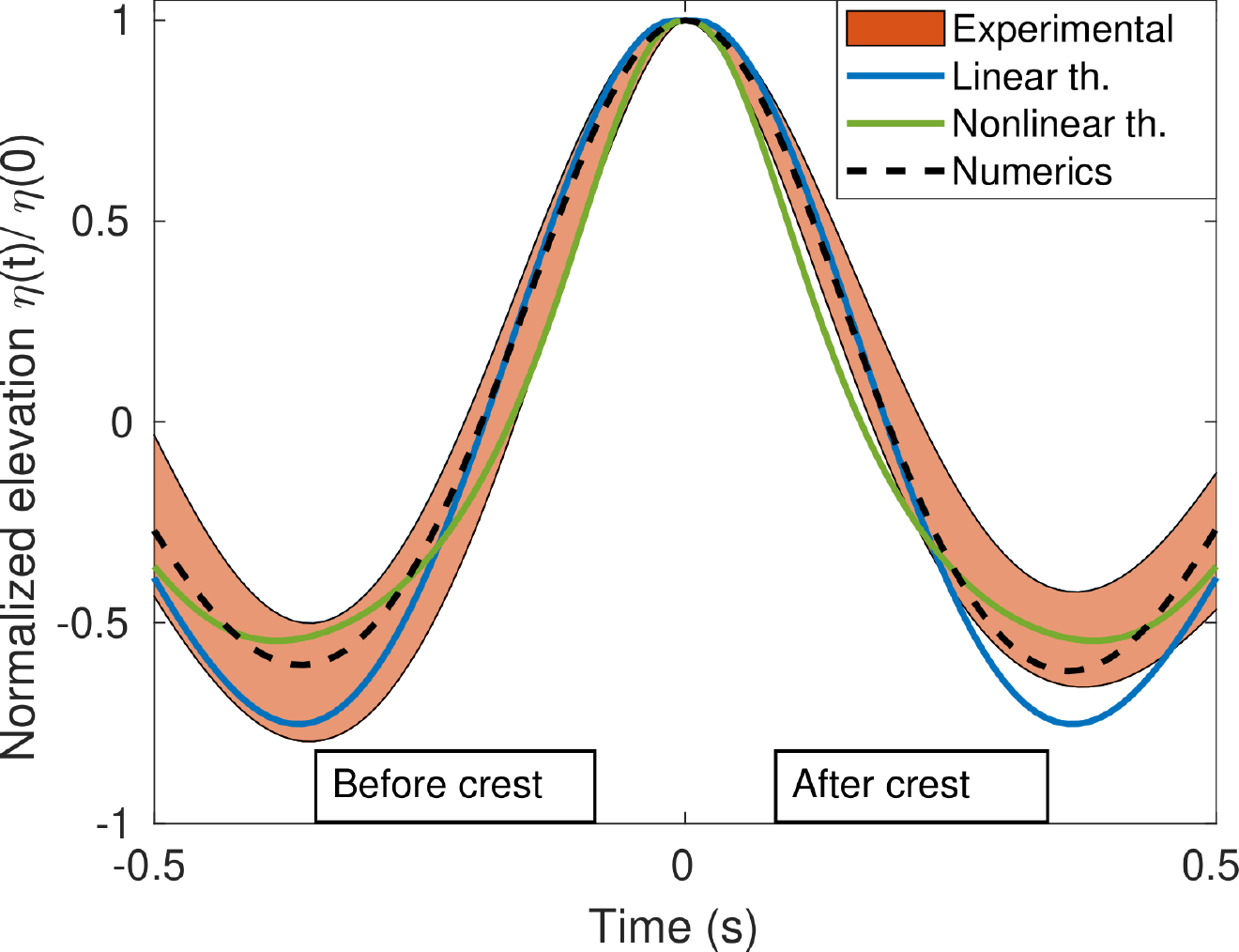} \end{center}

    \caption{Shape of the large crests ($\eta(0) > 1.25 H_S$) for Run 2. The coloured area corresponds to experiments (mean value $\pm$ standard deviation), the black dashed line to numerical models and solid lines to first- and second-order theories. Similar figures for Run 1 and Run 3 are reported in Appendix \ref{appA}.
    }
   \label{RW_shape}
 \end{figure}
 
The mean surface elevation at a given position right before/after a large crest occurs (identified as $\eta(0)$ with time origin shifted such as the crest manifests at $t=0$) is approximated at second order in the joint limit of small amplitude and frequency bandwidth as
\begin{equation}
\eta(t) = \eta(0)\left[ \frac{\Psi(t) + \frac{\eta_C \mathcal{F}(t)}{H_s}}{1+ \frac{\eta_C \mathcal{F}(0)}{H_s}}\right] ,\label{RW_profile}
\end{equation}
where $\Psi(t) = \langle \eta(0) \eta(t) \rangle / \sigma^2$ is the autocorrelation function, $\mathcal{F}$ is a function of $S_\eta$ detailed in Appendix \ref{appF}  and $\eta_C$ is the linear component of $\eta(0)$ \citep{Fedele_2009}. Previous studies have only tested this result in the linear limit in which $\eta_C = 0$  \citep{Soquet_2005, Klahn_2021}. The normalized elevation $\eta(t)/\eta(0)$ computed from Eq. \eqref{RW_profile} with both $\eta_C = 0$ and $\eta_C = 1.25 H_s$ is reported in Fig. \ref{RW_shape}, along with experimental and numerical values for crests such that $\eta_C>1.25 H_S$. Our data confirm that the linear approximation overestimates the depths of the troughs preceding and following the crest, a discrepancy fixed with second-order corrections. However, both theoretical models are symmetric in time reversal (since $\Psi(t) = \Psi(-t)$ and $\mathcal{F}(t) = \mathcal{F}(-t)$) whereas experimental measurements before and after the crest occurs persistently differ. This asymmetry also manifests in steeper slopes before the crests $(t<0)$ than after $(t>0)$. The numerical simulations of \cite{Fujimoto_2019} have shown that, at a fixed time and for directional wave fields, high crests are not symmetric in space as a result of the four-wave resonant interactions not captured by the second-order model reported in Eq. \eqref{RW_profile}.

\section{Conclusion}

Laboratory experiments with simplified directional spectra provide useful hints about the various processes taking place in the ocean without the usual bias of, e.g., wave breaking regularization in numerical simulations or varying environmental conditions in field measurements. In this study, more than two thousands RWs were observed in statistically homogeneous, isotropic and steady wave fields, allowing the predictions of commonly used theoretical models to be confronted with data in which strongly nonlinear events take place. To highlight the consequences of these high-order nonlinearities, numerical simulations associated with similar PSDs and valid up to second order were carried out. Therefore, they include the leading-order bound wave correction but not the resonant interactions.

The third and fourth normalized moments of surface elevation are compared with theoretical results in which the leading-order bound wave correction is accounted for. These analytic expressions are found to accurately describe the skewness of both experimental and numerical data. However, they significantly underestimate the experimental kurtosis while being in agreement with the numerical ones, evidencing a first consequence of resonant interactions on the statistics. This discrepancy is also manifest in the tails of the normalized surface elevation p.d.f.s.

The surface elevation time series are then split into individual waves whose heights, crests and troughs are analysed. The wave height is found to be robust to high-order effects, the experimental p.d.f.s being similar to the numerical ones and to the Rayleigh distribution. A similar conclusion cannot be drawn regarding the wave crests and troughs, for which large values are much more likely experimentally than numerically, indicating that four-wave resonant interactions strongly affect their statistics. The impact of high-order nonlinearities on large crests is further evidenced through the comparison of their mean shape with first- and second-order theoretical predictions, none of them being able to capture the asymmetry under time reversal. Therefore, the phenomenology of rogue waves crucially depend on how they are defined: high-order nonlinear effects do not seem to play a significant role if the wave height criterion $H> 8 \sigma$ is used, whereas for RW depicted as $\eta_\mathrm{C}> 5 \sigma$ (referred to as `large crests' in this paper) they significantly enhance their probability of occurrence. This finding demonstrates the current need for higher-order theoretical models that disentangle troughs and crests.

Note that, as reported in previous studies \citep{Aubourg_2017, Falcon_2020, Soquet_2005, Denissenko_2007, Klahn_2021}, some features of our second-order numerical model of \textit{isotropic} waves are surprisingly well fitted by theoretical models derived for \textit{unidirectional} and narrowband wave fields, provided that the single parameter they depend on, the steepness $\varepsilon_T$, is tuned to generate the observed skewness. This applies to the tails of the PDFs of both the normalized surface elevation and wave crests, but not to the wave troughs.

Many geophysical processes that are both challenging to model theoretically and to disentangle from other effects in field experiments could benefit from similar investigations with these isotropic nonlinear steady states. This includes, but is not limited to, the impact of waves on mixing and air-sea fluxes, the effect of rain in calming the sea and the effective parameters of random nonlinear waves (diffusion of a pollutant, damping and scattering of a wave train, etc.).

We thank the technical team at the ECN facilities for their help and support on the experimental setup. Part of this work was supported by the French National Research Agency (ANR DYSTURB Project No. ANR-17-CE30-0004), and by a grant from the Simons Foundation MPS No. 651463-Wave Turbulence. 

Declaration of Interests: The authors report no conflict of interest.

\appendix

\section{Detail on equation \eqref{S_th}}\label{appB}
Following \citet{Janssen_2009} 
and its notations, the third moment of the surface elevation $\mu_3$ is related to the standard deviation $\sqrt{\mu_2}$ and to the skewness parameter $C_3$ through its Eqs. (51) and (52), that are
\begin{equation}
C_3 = \frac{\mu_3}{\mu_2^{3/2}} = \frac{3}{m_0^{3/2}} \int \mathrm{d}\mathbf{k}_{1,2} E_1 E_2 \left( \mathcal{A}_{1,2} + \mathcal{B}_{1,2} \right),\label{SK1}
\end{equation}
where $m_0 = \int \mathrm{d}\mathbf{k}_1 E_1$ and $E(\mathbf{k})$ is the first-order spectrum. After lengthy but straightforward computations using various equations of \citet{Janssen_2009}
, we obtain the transfer coefficients $\mathcal{A}_{1,2} (\mathbf{k}_1,\mathbf{k}_2)$ and $\mathcal{B}_{1,2} (\mathbf{k}_1,\mathbf{k}_2)$ for deep-water gravity waves
\begin{align}
\mathcal{A}_{1,2}&= \frac{1}{\sqrt{k_1k_2}} \left[ \frac{\left(\sqrt{k_1}+\sqrt{k}_2\right)^2(\mathbf{k}_{1} \cdot \mathbf{k}_2 -k_1k_2)}{(\sqrt{k_1} +\sqrt{k_2})^2-\vert \mathbf{k}_1+ \mathbf{k}_2 \vert}-  \left(\frac{\mathbf{k}_{1} \cdot \mathbf{k}_2 - k_1k_2 - \sqrt{k_1k_2}(k_1+k_2)}{2}\right)    \right],\\
\mathcal{B}_{1,2} &= \frac{1}{\sqrt{k_1k_2}}\left[           \frac{(\sqrt{k_1}- \sqrt{k_2})^2 (\mathbf{k}_1\cdot \mathbf{k}_2 +k_1k_2)}{  (\sqrt{k_1}- \sqrt{k_2})^2- \vert \mathbf{k}_1 - \mathbf{k}_2 \vert} - \left( \frac{ \mathbf{k}_{1} \cdot \mathbf{k}_2 + k_{1}k_2- \sqrt{k_1k_2}(k_1+k_2)  }{2} \right)     \right].
\end{align}
It can be readily confirmed that this expression of the skewness corresponds to the one initially derived by Longuet-Higgins (Eq. (3.11) of \citet{LH_1963} 
corrected by a misprint of one half). Given that the wave field is assumed isotropic, $E(\mathbf{k}) \mathrm{d}\mathbf{k}= S_\eta(k)/(2 \pi) \mathrm{d}k \mathrm{d}\theta $, with $S_\eta(k)$ the surface elevation PSD. Moreover, since the transfer coefficients are invariant by a simultaneous rotation of $\mathbf{k}_1$ and $\mathbf{k}_2$, eq. \eqref{SK1} reduces to 
\begin{align}\label{SK2}
C_3 = & \frac{3}{m_0^{3/2}}\int_0^{2\pi} \mathrm{d} \theta \iint \mathrm{d}k_1\mathrm{d}k_2 \frac{S_\eta(k_1) S_\eta(k_2)}{2 \pi \sqrt{k_1k_2}}\\
&\left[ \frac{\left(\sqrt{k_1}+\sqrt{k}_2\right)^2(\mathbf{k}_{1} \cdot \mathbf{k}_2 -k_1k_2)}{(\sqrt{k_1} +\sqrt{k_2})^2-\vert \mathbf{k}_1+ \mathbf{k}_2 \vert}+ \frac{(\sqrt{k_1}- \sqrt{k_2})^2 (\mathbf{k}_1\cdot \mathbf{k}_2 +k_1k_2)}{  (\sqrt{k_1}- \sqrt{k_2})^2- \vert \mathbf{k}_1 - \mathbf{k}_2 \vert} - \mathbf{k}_{1} \cdot \mathbf{k}_2 + \sqrt{k_1k_2}(k_1+k_2)    \right]\nonumber, 
\end{align}
with $\mathbf{k}_1 = k_1 \mathbf{e}_x$ and $\mathbf{k}_2 = k_2 \left( \cos \theta \mathbf{e}_x + \sin \theta \mathbf{e}_y \right)$. Define a function $I$ such that 
\begin{equation}
I(\alpha) = \int_0^{2\pi} \mathrm{d} \theta \left[  \frac{\sqrt{\alpha}\left(1+\sqrt{\alpha}\right)^2(\cos \theta - 1)}{(1 +\sqrt{\alpha})^2-\sqrt{1+\alpha^2 + 2 \alpha \cos \theta}}+  \frac{\sqrt{\alpha}(1- \sqrt{\alpha})^2 (\cos \theta + 1)}{  (1- \sqrt{\alpha})^2- \sqrt{1 + \alpha^2 - 2 \alpha \cos \theta}}  + (1+\alpha)    \right],
\end{equation}
and \eqref{SK2} then reads
\begin{equation}
C_3 = \frac{3}{m_0^{3/2}}\iint  \frac{S_\eta(k_1) S_\eta(k_2) k_1}{2 \pi} I\left( \frac{k_2}{k_1} \right)\mathrm{d}k_1\mathrm{d}k_2,
\end{equation}
which corresponds, with $C_3 \rightarrow \mathcal{S}$ and $m_0 \rightarrow \sigma^2$ (our notations), to Eq. \eqref{S_th}.

\section{Detail on equation \eqref{C_th}}\label{appC}
A similar procedure can be applied to compute the canonical contribution to the kurtosis from Eq. (59) of \citet{Janssen_2009}
,
\begin{equation}
C_4^\mathrm{can} = \frac{4}{m_0^2} \int  E(\mathbf{k}_1)E(\mathbf{k}_2)E(\mathbf{k}_3)\Psi(\mathbf{k}_1,\mathbf{k}_2,\mathbf{k}_3) \mathrm{d}\mathbf{k}_1\mathrm{d}\mathbf{k}_2\mathrm{d}\mathbf{k}_3,
\end{equation}
where $\Psi$ is an explicit interaction coefficient not detailed here. With $E(\mathbf{k}_i) = S_\eta(k_i)/(2\pi) \mathrm{d}k_i \mathrm{d}\theta_i$ and $\sigma^2 = m_0$,  
\begin{equation}
C_4^\mathrm{can} = \frac{4}{(2\pi)^3\sigma^4} \int  S_\eta(k_1)S_\eta(k_2)S_\eta(k_3)\Psi(\mathbf{k}_1,\mathbf{k}_2,\mathbf{k}_3) \mathrm{d}k_1\mathrm{d}k_2\mathrm{d}k_3\mathrm{d}\theta_1\mathrm{d}\theta_2\mathrm{d}\theta_3.
\end{equation}
Since $\Psi$ is invariant under a simultaneous rotation of $\mathbf{k}_1$, $\mathbf{k}_2$ and $\mathbf{k}_3$, a first integration can be performed
\begin{equation}
C_4^\mathrm{can} = \frac{4}{(2\pi)^2\sigma^4} \int  S_\eta(k_1)S_\eta(k_2)S_\eta(k_3)\Psi(k_1 \mathbf{e}_x,\mathbf{k}_2,\mathbf{k}_3) \mathrm{d}k_1\mathrm{d}k_2\mathrm{d}k_3\mathrm{d}\theta_2\mathrm{d}\theta_3,
\end{equation}
with $\mathbf{k}_{2,3}= k_{2,3} \left( \cos \theta_{2,3} \mathbf{e}_x + \sin \theta_{2,3} \mathbf{e}_y \right)$. Finally, note that the function $\Psi$ is such that 
\begin{equation}
\Psi(k_1 \mathbf{e}_x,\mathbf{k}_2,\mathbf{k}_3) = k_1^2 \Psi\left(\mathbf{e}_x,\frac{\mathbf{k}_2}{k_1},\frac{\mathbf{k}_3}{k_1}\right),
\end{equation}
and define a function $\psi$ by
\begin{equation}
\psi\left(\alpha, \beta \right) = \int \Psi\left( \mathbf{e}_x, \alpha \left[ \cos \theta_2 \mathbf{e}_x + \sin \theta_2 \mathbf{e}_y \right], \beta \left[ \cos \theta_3 \mathbf{e}_x + \sin \theta_3 \mathbf{e}_y \right] \right) \mathrm{d}\theta_2 \mathrm{d}\theta_3.
\end{equation}
The coefficient $C_4^\mathrm{can}$ then reads
\begin{equation}
C_4^\mathrm{can} = \frac{4}{(2\pi)^2\sigma^4} \int  S_\eta(k_1)S_\eta(k_2)S_\eta(k_3)k_1^2 \psi\left(\frac{k_2}{k_1},\frac{k_3}{k_1}\right)  \mathrm{d}k_1\mathrm{d}k_2\mathrm{d}k_3,
\end{equation}
which corresponds to Eq. \eqref{C_th}.

\section{Tayfun p.d.f. of surface elevation} \label{appD}
The p.d.f. of surface elevation can be explicitly computed in the case of a unidirectional and narrowband wave field in which only the first nonlinear correction is computed. However, several misprints make the expression of this PDF difficult to obtain from the literature. In particular, the original derivation of \citet{Tayfun_1980} 
must be corrected as follows: his Eq. (24) should read
\begin{equation}
F_\xi (u) = \left( 2\pi\right)^{\textcolor{red}{-1/2}} \int_{\alpha(u)}^\infty  e^{-\tau^2/2}\left\lbrace \mathrm{erf}\left[A(\tau,u) + \beta \right]+\mathrm{erf}\left[A(\tau,u) - \beta \right] \right\rbrace  d\tau,
\end{equation}
and his corrected Eq. (27) is 
\begin{equation}
A(\tau, u) = \beta \sqrt{1+\frac{\sqrt{2}\gamma u}{\beta}+ \frac{\tau^2}{\textcolor{red}{2}\beta^2}}.
\end{equation}
Note also that only approximate expressions of this p.d.f. are reported in \citet{Soquet_2005}
: indeed, their Eq. (7) becomes undefined for large negative values of the surface elevation (if their $1+2\sigma z <0$, their $C(0)$ required in the integral is no longer real valued).

For completeness, the full set of equations required to compute the p.d.f. $f(u)$ of the normalized surface elevation $u = \eta/\sigma$ ($\sigma = \langle \eta^2 \rangle^{1/2}$) is reported below
\begin{equation}
f(u) = \frac{\mathrm{d} F}{\mathrm{d}u}, ~~~F(u) = \frac{1}{\sqrt{2\pi}} \int_{\alpha(u)}^\infty e^{- \frac{\tau^2}{2}} \left[ \mathrm{erf}\left(A(\tau,u)+ \beta \right) +  \mathrm{erf}\left(A(\tau,u)- \beta \right)  \right]  \mathrm{d}\tau
\end{equation}
with
\begin{equation}
A(\tau,u) = \beta \sqrt{1+ \frac{\sqrt{2}\gamma u}{\beta} + \frac{\tau^2}{2\beta^2}}, ~\beta = \frac{1}{\sqrt{-1 + \sqrt{1+ 4 \sigma^2 k^2}}}, ~\gamma = \sqrt{\frac{1+ \sqrt{1+4 \sigma^2 k^2}}{2} },
\end{equation}
and
\begin{equation}
\alpha \left(u \geqslant - \frac{\beta}{\sqrt{2}\gamma} \right) = 0, ~~~\alpha \left(u < - \frac{\beta}{\sqrt{2}\gamma} \right) = \beta\sqrt{-2\left(1+ \frac{\sqrt{2}\gamma u}{\beta}\right)}.
\end{equation}

\section{Second- and third-order Gram-Charlier series} \label{appD2}
A theoretical approach to the p.d.f. of surface elevation consists in using low-order Gram-Charlier series. Following \cite{Klahn_2021}, we define in this manuscript the second-order approximation as 
\begin{equation}
f_{GC2}\left(u = \frac{\eta}{\sigma}\right) = \frac{1}{\sqrt{2\pi}}e^{- u^2/2} \left[ 1 + \frac{\mathcal{S}}{6}H_3(u)  \right], ~~~ H_3(u) = u^3 - 3u,
\end{equation}
and the third-order one
\begin{equation}
f_{GC3}\left(u = \frac{\eta}{\sigma}\right) = \frac{1}{\sqrt{2\pi}}e^{- u^2/2} \left[ 1 + \frac{\mathcal{S}}{6}H_3(u)  + \frac{1}{24}(K-3) H_4(u) + \frac{1}{72}\mathcal{S}^2 H_6(u)   \right], 
\end{equation}
with 
\begin{equation}
H_4(u) = u^4 - 6u^2+3, ~~ H_6(u) = u^6 - 15u^4 + 45u^2 - 15.
\end{equation}

\section{Tayfun p.d.f. of the crests and troughs} \label{appE}
For unidirectional and narrowband wave fields, the p.d.f. of crests accounting for second-order nonlinearities reads \citep{Tayfun_1980}
\begin{equation}
f_C(\xi_\mathrm{C}) = \frac{2 \varepsilon}{-1 + \sqrt{1 + 4 \varepsilon^2}} \left(1 - \frac{1}{\sqrt{1+ 2 \varepsilon \xi_\mathrm{C}}} \right)\exp \left[- \frac{\left(-1+\sqrt{1+2 \varepsilon \xi_\mathrm{C}} \right)^2}{-1+\sqrt{1+4 \varepsilon^2}}   \right],
\end{equation}
with $\xi_\mathrm{C} = \eta_\mathrm{C} / \sigma$ and $\sigma = \langle \eta^2 \rangle^{1/2}$ (note that the PDF reported in \citet{Tayfun_1980} 
considers instead the wave crest normalized by the standard deviation of the linear component). The steepness parameter $\varepsilon = \sigma k$, with $k$ the central wavenumber of the narrowband wave fields, is in that case related to the skewness $\mathcal{S}= 3 \varepsilon + O(\varepsilon^3)$. Similarly, for the troughs,
\begin{equation}
f_T(\xi_\mathrm{T}) = \frac{-2 \varepsilon}{-1 + \sqrt{1 + 4 \varepsilon^2}} \left(1 - \frac{1}{\sqrt{1- 2 \varepsilon \xi_\mathrm{T}}} \right)\exp \left[- \frac{\left(-1+\sqrt{1+2 \varepsilon \xi_\mathrm{T}} \right)^2}{-1+\sqrt{1+4 \varepsilon^2}}   \right],
\end{equation}  
with $\xi_\mathrm{T} = \eta_\mathrm{T} / \sigma$.

\section{Additional wave features}\label{appA}
\begin{figure}
    \begin{center} 
    \includegraphics[height=4.70cm]{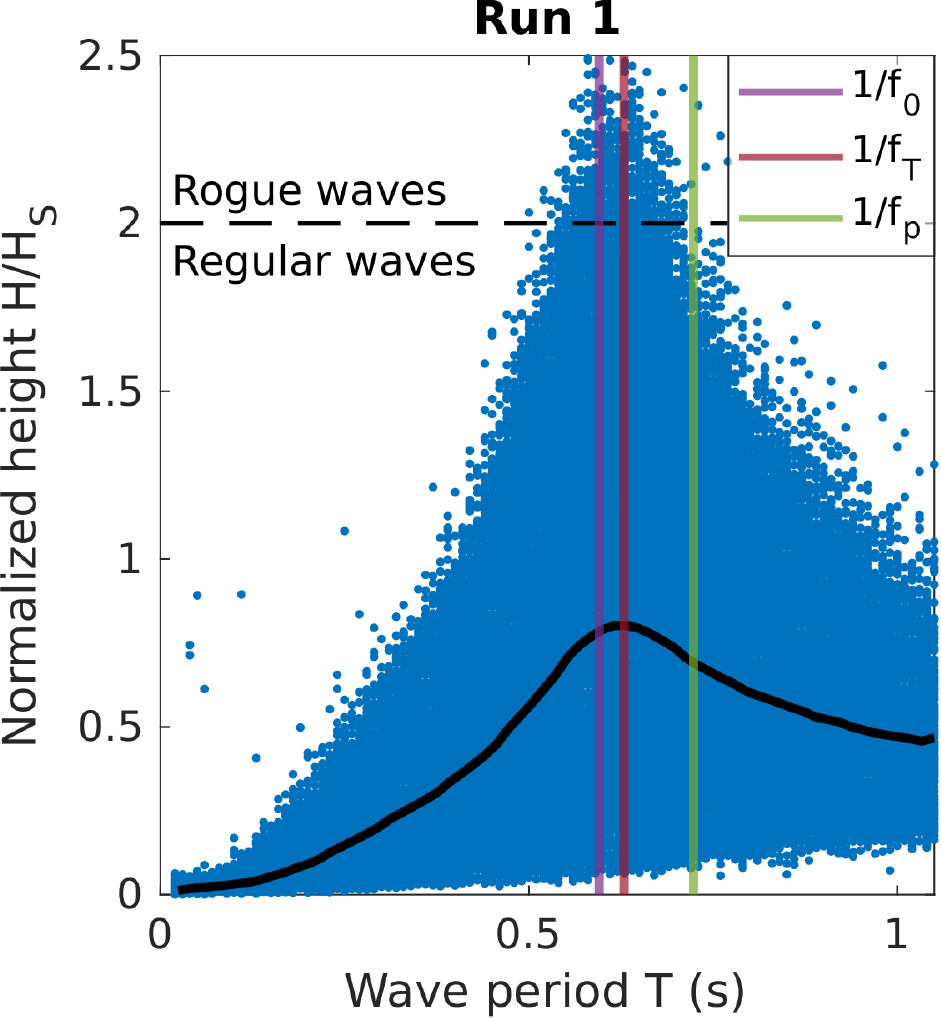}
    \includegraphics[height=4.70cm]{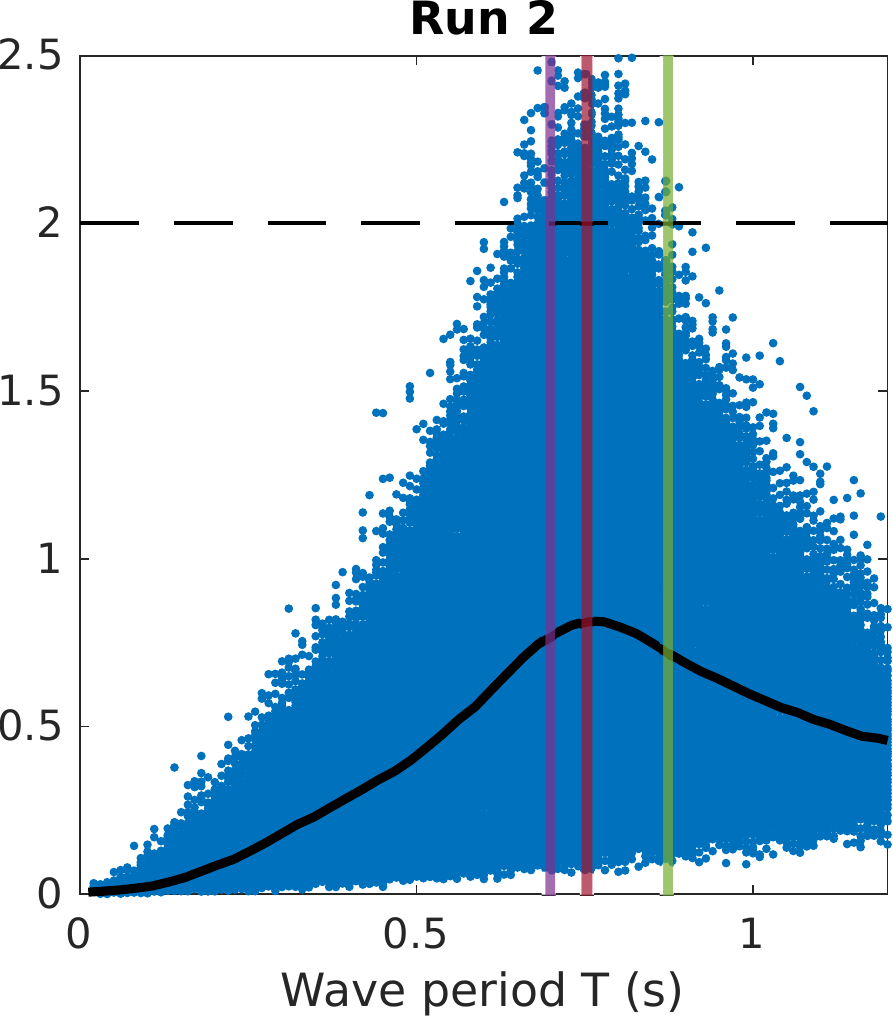}
    \includegraphics[height=4.70cm]{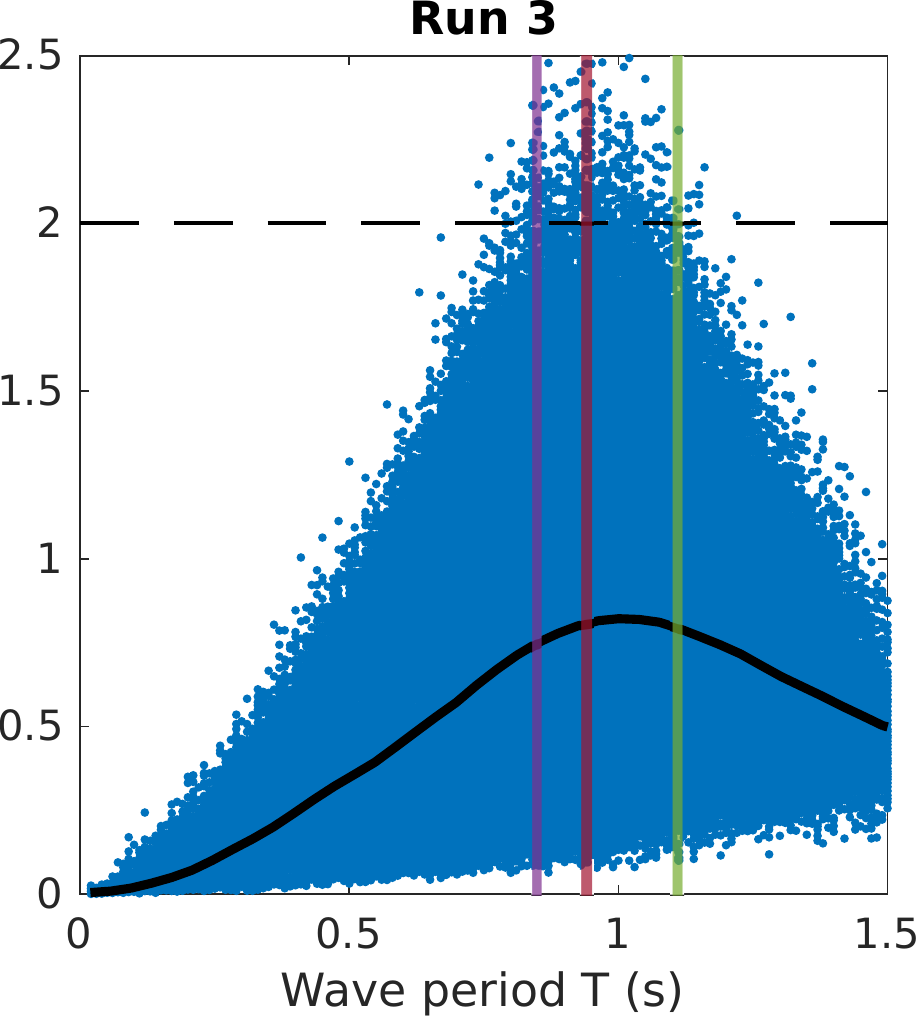}
    \caption{Experimental normalized wave height as a function of the wave period. The mean value is plotted in thick black and peaks close to the Tayfun period $1/f_T$. Vertical lines indicate $f_0^{-1}$, $f_T^{-1}$ and $f_\mathrm{p}^{-1}$ ($f_0^{-1}<f_T^{-1}<f_\mathrm{p}^{-1})$}
    \label{H_vs_T}
\end{center}
\end{figure}

\begin{figure}
    \begin{center} 
    \includegraphics[height=3.75cm]{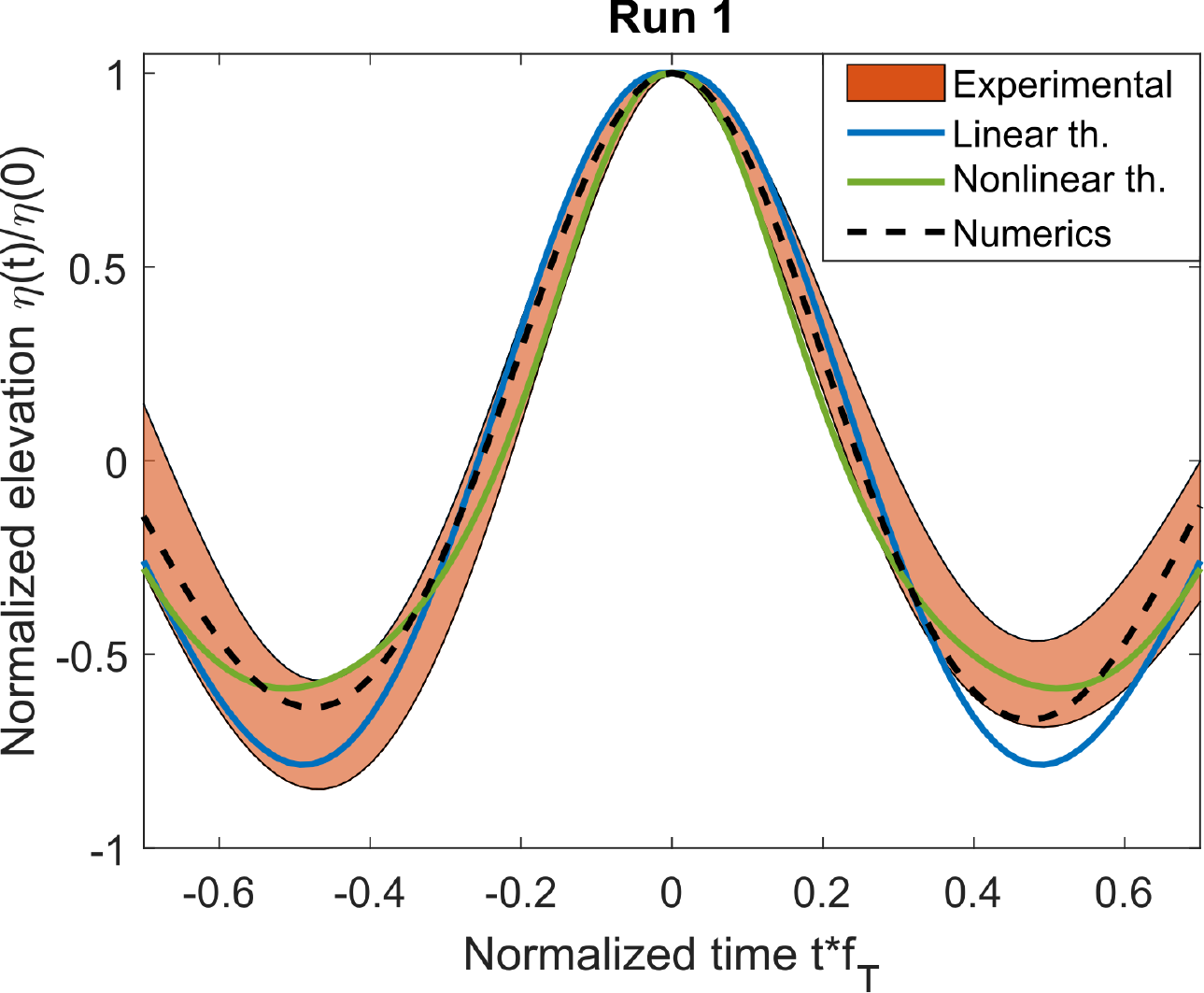}
    \includegraphics[height=3.75cm]{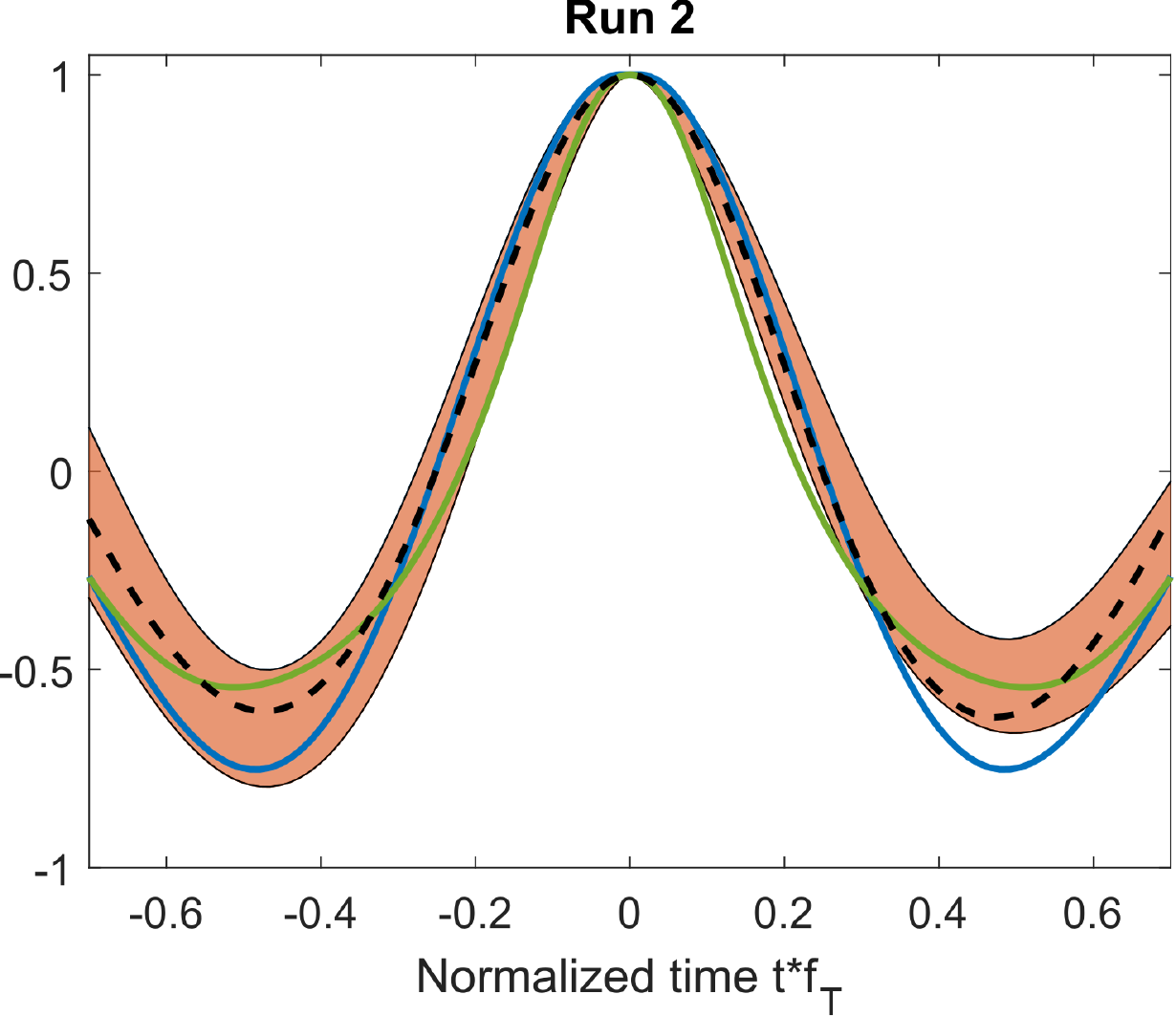}
    \includegraphics[height=3.75cm]{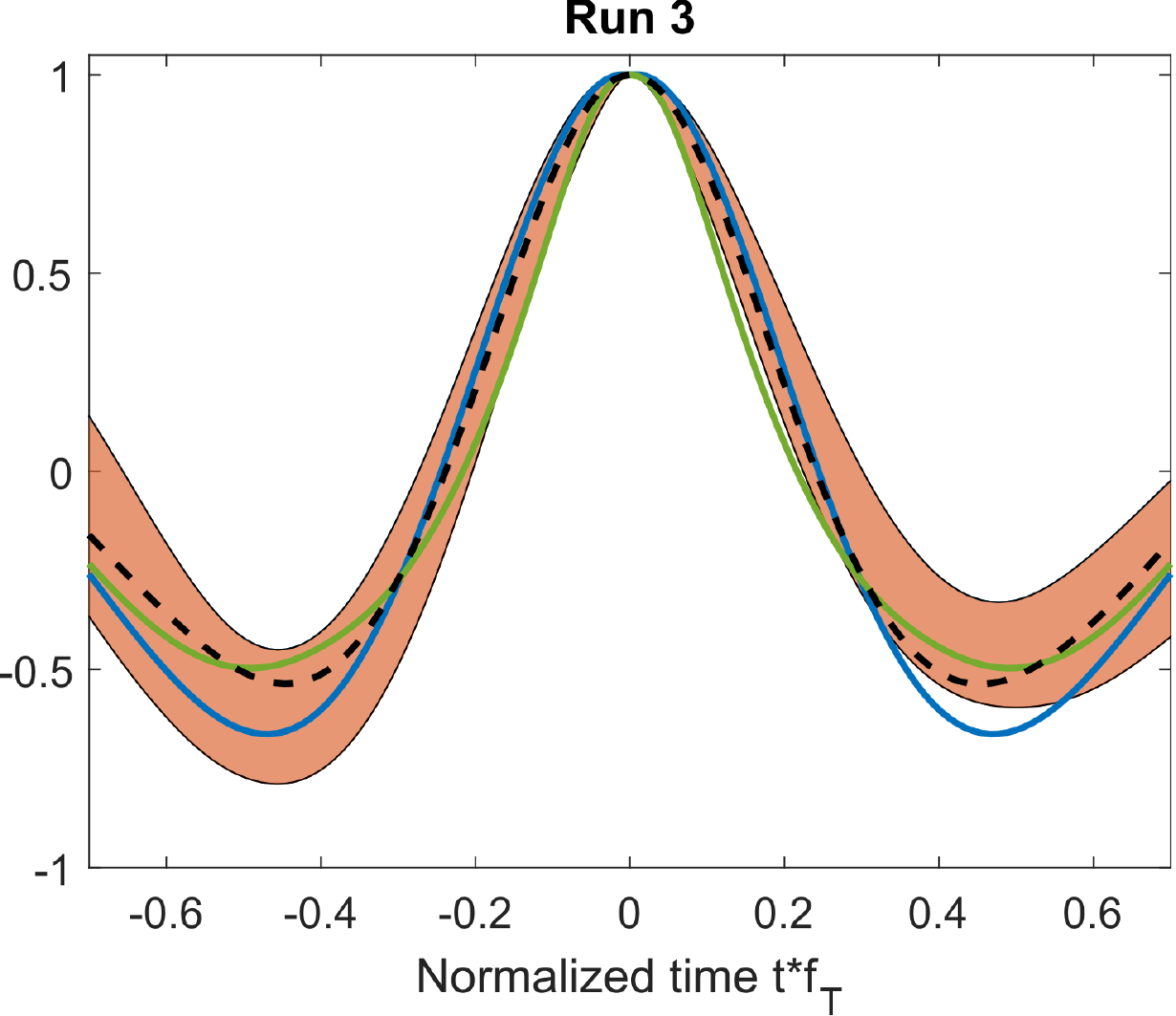}
    \caption{Shape of large crests $(\eta(0)> 1.25 H_S)$ for Runs 1, 2 and 3. The coloured area corresponds to experiments (mean value $\pm$ standard deviation), the black dashed line to numerical models and solid lines to first- and second-order theories. The central figure corresponds to Fig. \ref{RW_shape}.}
    \label{Large_waves}
\end{center}
\end{figure}

The raw data of the normalized wave height $H/H_S$ plotted versus the wave period $T$ are reported in Fig. \ref{H_vs_T}, while the shape of the large crests for Runs 1, 2 and 3 are shown in Fig. \ref{Large_waves}

\section{Expected shape of large waves}\label{appF}
From Eq. (5.7) of \citet{Fedele_2009}
, define in the deep-water and isotropic limit the function
\begin{align} \nonumber
\mathcal{F} (t)&= \frac{2}{(2\pi)^2\sigma^3} \int S_\eta(k_1) S_\eta(k_2) \\&\left[ (\mathcal{A}_{1,2} + \mathcal{B}_{1,2}) \cos(\omega_1t) \cos(\omega_2t) - (\mathcal{A}_{1,2} - \mathcal{B}_{1,2}) \sin(\omega_1t) \sin(\omega_2t)\right] dk_1 dk_2 d\theta_1 d\theta_2,
\end{align}
with $\omega_{1,2} = \sqrt{g k_{1,2}}$. A first angular integration can be performed to obtain
\begin{align} \nonumber
\mathcal{F} (t)&= \frac{1}{\pi\sigma^3} \int S_\eta(k_1) S_\eta(k_2)\\& \left[ (\mathcal{A}_{1,2} + \mathcal{B}_{1,2}) \cos(\omega_1t) \cos(\omega_2t) - (\mathcal{A}_{1,2} - \mathcal{B}_{1,2}) \sin(\omega_1t) \sin(\omega_2t)\right] dk_1 dk_2 d\theta,
\end{align}
with $\mathbf{k}_1 = k_1 \mathbf{e}_x$ and $\mathbf{k}_2 = k_2 \left( \cos \theta \mathbf{e}_x + \sin \theta \mathbf{e}_y \right)$. Further assume
\begin{equation}
J(\alpha) = \sqrt{\alpha} \int_0^{2\pi} \mathrm{d} \theta \left[  \frac{\left(1+\sqrt{\alpha}\right)^2(\cos \theta - 1)}{(1 +\sqrt{\alpha})^2-\sqrt{1+\alpha^2 + 2 \alpha \cos \theta}}-  \frac{(1- \sqrt{\alpha})^2 (\cos \theta + 1)}{  (1- \sqrt{\alpha})^2- \sqrt{1 + \alpha^2 - 2 \alpha \cos \theta}}  + 2    \right],
\end{equation}
to obtain
\begin{equation}
\mathcal{F}(t) = \frac{1}{\pi \sigma^3} \int S_\eta(k_1) S_\eta(k_2) k_1 \left[ I\left( \frac{k_2}{k_1}\right) \cos(\omega_1t) \cos(\omega_2t) - J\left( \frac{k_2}{k_1}\right) \sin(\omega_1 t) \sin(\omega_2t) \right] \mathrm{d}k_1 \mathrm{d}k_2
\end{equation}
which allows simple numerical integration. The elevation profile $\eta(t)$ close to a crest of linear elevation $\xi_c$ then follows from Eq. (5.10) of \citet{Fedele_2009} 
and reads at leading order
\begin{equation}
\eta(t) =  \xi_c \Psi(t) + \frac{\xi_c^2 \mathcal{F}(t)}{4\sigma},
\end{equation}
with $\Psi(t) = \langle \eta(0) \eta(t) \rangle / \sigma^2$ the autocorrelation function. \\

\bibliographystyle{jfm}

\end{document}